# Parallel hardware for faster morphological analysis

Issam Damaj [a,*], Mahmoud Imdoukh [a], Rached Zantout [b]

[a] Department of Electrical and Computer Engineering, American University of Kuwait, P.O. Box 3323, Safat 13034, Kuwait
[b] Department of Electrical and Computer Engineering, Rafik Hariri University, P.O. Box 10, Mechref, Damour, Chouf, 2010, Lebanon

ABSTRACT

Morphological analysis of Arabic language is computationally intensive, has numerous forms and rules, and intrinsically parallel. The investigation presented in this paper confirms that the effective development of parallel algorithms and the derivation of corresponding processors in hardware enable implementations with appealing performance characteristics. The presented developments of parallel hardware comprise the application of a variety of algorithm modelling techniques, strategies for concurrent processing, and the creation of pioneering hardware implementations that target modern programmable devices. The investigation includes the creation of a linguistic-based stemmer for Arabic verb root extraction with extended infix processing to attain high-levels of accuracy. The implementations comprise three versions, namely, software, non-pipelined processor, and pipelined processor with high throughput. The targeted systems are high-performance multi-core processors for software implementations and high-end Field Programmable Gate Array systems for hardware implementations. The investigation includes a thorough evaluation of the methodology, and performance and accuracy analyses of the developed software and hardware implementations. The developed processors achieved significant speedups over the software implementation. The developed stemmer for verb root extraction with infix processing attained accuracies of 87% and 90.7% for analyzing the texts of the *Holy Quran* and its Chapter 29 – *Surat Al-Ankabut*.

## 1. Introduction

Natural Language Processing (NLP) is a rapidly developing field. Developments in NLP are, at the larger part, driven by the fact that the world has turned into a small village equipped with advanced transportation, media, and communication. In 2016, the total number of social network users worldwide is estimated to be around 2.2 billion with a global penetration of 31%. In the US, 78% of the population has social network profiles. Indeed, it is expected that the total number of users will grow to 2.5 billion in 2018 (Statista, 2017). In such a modern, connected, and global society, people still use different languages. The idea of having all the people use one language has been proven by practice to be impossible. Even in science fiction, as in Star Trek, a universal translator is presented to solve the challenge of automatic translation among languages. At present, NLP active areas of research are machine translation, information retrieval, text categorization, sentiment mining, to name a few (Nirenburg and Wilks, 2000; Yang et al., 2012; Agarwal and Mittal, 2016).

A Morphological Analyzer (MA) is a core subsystem in NLP applications. MAs work on identifying words being used in a specific language and study the internal structure of these words (Hamalawy, 2009). Morphology can be defined as producing a word from another by changing it to fit a new meaning. Furthermore, Morphological analysis is usually complicated, computationally intensive, and intrinsically parallel. Arabic language is well-known for having rich morphology, complex word formation and patterns (Al-Sughaiyer and Al-Kharashi, 2004).

### 1.1. Background

The rich morphology of Arabic enables the language to develop and grow. Arabic morphology (الإشتقاق) is categorized into small, large, larger, and the largest morphologies. The small morphology derives a word from a root but keeps similarities between the two words in their pronunciation and meaning; such as (علم ,عالم) that translates to (root: science, derived: scientist). The remaining morphologies comprise exchanging letters of the root, producing a word from another by changing one or more letters, and producing a word by combining a group of words. By far, the mostly used type of morphology in Arabic is the small morphology (Al-Sughaiyer and Al-Kharashi, 2004; Soudi et al., 2007; Dahdah, 1995; Rajhi, 1979; Hamandi et al., 2006; Al-Khalifah, 1996).

* Corresponding author.
*E-mail addresses:* idamaj@auk.edu.kw (I. Damaj), s00024916@alumni.auk.edu.kw (M. Imdoukh), Zantoutrn@rhu.edu.lb (R. Zantout).

Arabic words are grouped into the three main categories of Nouns, Verbs, and Particles. The Nouns and Verbs consist of subcategories in which the main form of the word changes per its position/role in the sentence and some of the other words in the sentence. For example, verbs are categorized per time and structure. The verb times are past, present or future, while structures may be either proper or defective. The difference between words in subcategories can be as subtle as the presence/absence of a vowel or as clear as the addition/removal of letters to/from the word.

In Arabic language, small morphology can act on verbs (Al-Sughaiyer and Al-Kharashi, 2004). The roots of Arabic words have traditionally been considered to consist of three or four letters. Like other languages, in Arabic, letters can be added to the beginning (prefixes) and/or to the end (suffixes) of the root. However, in Arabic, letters can be added inside a root (infixes). Infixes complicate the morphological analysis of the Arabic language because the infix letters can also be letters in the root. In addition, a Verb in Arabic has different forms if the subject or object is masculine or feminine. Also, differences in the forms can exist if the sentence refers to one person, a group of two, or a group of more than two (Al-Khalifah, 1996). Nevertheless, Arabic verbs follow specific forms. For example, the root (درس, Study) maps to the ternary pattern (فعل), while the verb (يدرس) maps to (يفعل). Here, the addition of the prefix (ي) derives the present tense of the verb.

In Arabic verbs, there are seven letters that can be added to the beginning of the root as prefixes; these letters are grouped in the Arabic word (فسألتني). The nine letters that can be added to the end of the root as suffixes are grouped in the Arabic word (ايتهكمون). The five letters that can be added to the inside of a root (infixes) are grouped in the Arabic word (اتوني); infixes have a more complicated set of rules with focus on the three vowel letters و ,ا, and ي (Hamandi et al., 2006). In Table 1, example morphological variations from a verb root is presented with focus on the applied change on form and meaning. The same patterns shown in Table 1 can produce similar variations for the verb root (صحب, Accompany) to produce يصحبون and يصاحب in the same tense and form as in the patterns يفعلون and يفاعل. As compared to verbs, Arabic nouns are more complex due to the large variety of forms, irregularities, and number.

### 1.2. Related work

Al-Sughaiyer and Al-Kharashi (2004) presented a comprehensive survey of Arabic morphological analysis techniques that comprises definitions, classifications, approaches, algorithms, etc. In the literature, different MA techniques and algorithms target Arabic verbs specifically. Yaghi et al. (2003) presented a verb generation system that enables word-to-root and root-to-word lookups. The system uses coding techniques to compactly store and efficiently access a dictionary of Arabic words. Yagi and Harous (2003) details the development of a database of generated stems that supports their verb-matching system. The developed database is of multipurpose and can be used to identify stems, classify morpho-semantic and morpho-syntactic templates, and support a variety of applications. Boubas et al. presented the use of genetic algorithms to generate an MA for Arabic verb. The investigations comprised developing general verb patterns and then applying them to derive morphological rules. The reported results reflect highly accurate matching capabilities (Boubas et al., 2011).

A variety of algorithms have been developed to perform morphological analysis of verbs based on an input word, such as, sliding window algorithms (El-Affindi, 1998), word decomposition using algebraic algorithms (El-Affindi, 1991), and literals generation using permutations of the input word letters (Al-Shalabi and Evens, 1998). Other analyzers attempt to extract the root of a verb by manipulating infixes and prefixes (Hamandi et al., 2002, 2006; Khoja, 2017; Khoja and Garside, 1999; Larkey and Connell, 2006; Saad et al., 2010; Larkey et al., 2002; Asaad and Abbod, 2014; Boudlal et al., 2011; Hegaz and Elsharkawi, 1986; Hlal, 1987; Abu Shquier and Alhawiti, 2015; Sembok and Ata, 2013; Abu-Errub et al., 2014; Al-Bawab and Al-Tayyan, 1998). The extracted stems are then validated against a list of standard Arabic roots. In (Hamalawy, 2009), such a manipulation of affixes is classified under Linguistic-based (LB) stemmers. LB stemmers are usually accurate but require the preparation of lists for matching and validation. If a stemmer doesn't include analysis of infixes and root extraction, it is referred to as a light stemmer (Larkey et al., 2002).

LB stemmers attracted the attention of many researchers and enabled the development of a variety of MAs. The focus of the presented MAs is accuracy; however, almost all contributions highlight the essential need for high-performance processing. Khoja (2017) and Khoja and Garside (1999) presented the development of an LB Arabic MA algorithm; the algorithm analyzes a word by removing definite articles, prefixes, suffixes, stop words, and then matches the remaining word against the pattern of the same length to extract the root. Khoja stemmer is widely used in the literature with a reported accuracy of 96% (Khoja, 2017). Asaad and Abbod (2014) presented an extraction approach that removes prefixes, suffixes, infixes, and attempts to identify the root. The presented approach includes making a second attempt to identify an unidentified root through a procedure that handles weak, hamzated (that has the letter Hamza 'ء'), eliminated-long-vowels, and two-letter geminated words. The proposed approach produced somewhat improved accuracy over Khoja stemmer. Boudlal et al. (2011) presented an Arabic MA system that extracts roots depending on the context within a sentence. A Hidden Markov Models approach was used, where the observations are the words and the possible roots represent the hidden states. The approach achieved an accuracy of 94% in targeting the NEMLAR Arabic writing corpus with its 500,000 words. LB stemmers have a long history of reported contributions since 1985 (Al-Sughaiyer and Al-Kharashi, 2004); this includes the approaches of Hegaz and Elsharkawi (1986), Hlal (1987), Abu Shquier and Alhawiti (2015), Sembok and Ata (2013), Abu-Errub et al. (2014), El-Affindi (1998, 1991), Al-Bawab and Al-Tayyan (1998), Khoja (2017) and Khoja and Garside (1999), to name but a few. Indeed, all the reported contributions above are developed as software implementations.

**Table 1**
Morphological variations of the verb Study (درس) with changes on form and meaning.

| Addition | Location | Morph | Pattern | Meaning | (Tense, Form) |
|---|---|---|---|---|---|
| (ي) | Prefix | يدرس | يفعل | One is studying | (Present, Singular) |
| (ي, ون) | (Prefix, Suffix) | يدرسون | يفعلون | Many are studying | (Present, Plural) |
| (ي, ا) | (Prefix, Infix) | يدارس | يفاعل | One is studying with others | (Present, Singular) |

LB stemming can also be used for nouns, however, with higher complexity than verbs.

## 2. Research objectives

Challenges to Arabic MAs comprise performance characteristics (speed, efficiency, and complexity), storage requirements, reliability and accuracy, and first and foremost the dealing with the language morphology. Arabic language morphological structure leads to huge lexical variations and number of dissimilar forms. Many definite articles, conjunctions, and particles can be found as prefixes. The number of possible suffixes is astronomical. In addition, many stems are derived from roots by infixing. For example, Table 2 presents a larger set of morphological variations for the verb root Study (درس) than the example in Table 1; additional complexity arises from the use of the diacritics Fatha (َ), Kasra (ِ), Damma (ُ), Sukun (ْ), and Shadda (ّ). Table 2 is generated by Qutrub (2017) and shows 82 different forms that can be reduced to 36 without the diacritics. Even closely-related forms, such as singular and plural, are not related by simple affixing. Another example of Arabic MA complexity includes the need to deal with deleted letters from words. For instance, the root Saw (رأى) has the word See (يرى) in its present tense; the present tense is deviated from the standard pattern (يفعل) that would have given the wrong standard word of (يرأى). The letter “”أ is deleted from the word that follows the standard pattern to produce the correct word (يرى) which doesn't follow the standard form of present tense (Dahdah, 1995; Asaad and Abbod, 2014). In addition, MA algorithms are computationally intensive and demanding when it comes to computing resources.

Modern high-performance computers (HPCs) are hybrids of multi-core processors, graphical processing units (GPUs), high-density Field Programmable Gate Arrays devices (FPGAs), to name a few. Within hybrid systems, algorithms can be partitioned and distributed or fully-delegated to one co- or pre-processing subsystem. Hybrid HPCs are supported by rich co-analysis and co-design tools that enable unified hardware/software implementations and effective rapid prototyping of hardware to run computationally intensive algorithms (Damaj, 2006, 2007; Damaj and Diab, 2003; Kasbah et al., 2008; Damaj and Kasbah, 2017). With no doubt, MAs can benefit from state-of-the-art HPCs to run on software and/or hardware implementations with optimized performance characteristics.

Limited work has been reported in the literature to present hardware implementations of MA systems in general and for Arabic language in specific. Murty et al. (2003), and like many investigations, presented the design of high-speed string matching co-processor for NLP. Other reported work comprises fast VLSI implementations for approximate string matching (Grossi, 1992), FPGA-based co-processor for text extraction (Ratha et al., 2000), and ASIC design of a high-speed unit for NLP to match inputs with lexical entries (Raman and Shaji, 1995). Cohen (1998) presented a hardware-assisted algorithm for large-dictionary string matching using n-gram hashing. The proposed hardware system comprises a personal computer with a co-processing board. Moreover, Hamandi et al. (2006) stressed the importance of using parallel processing and hardware implementation to accelerate NLP and MA systems. The authors presented a sample parallel algorithm for an LB stemmer and reasoned about its characteristics. Nevertheless, the authors didn't include sample IP-cores or present implementation results. Parallel and hardware processing are common in information retrieval, speech recognition, text-to-speech, speech synthesis etc. (Rasmussen, 1991; Gadri and Moussaoui, 2015; Mahdaouy et al., 2014; Sensory, 2017; Khodor and Zaki, 2011). Ultimately, the aim of all presented hardware investigations is at making NLP systems more responsive by providing faster processing.

This paper presents high-speed hardware implementations of an Arabic LB stemmer under FPGAs; the targeted stemmer is for Arabic verb root extraction. The developed implementations carefully exploit the intrinsic parallelism in LB stemmers to provide an enhanced performance with adequate accuracy levels. The hardware intellectual property (IP)-cores are systematically developed using the methodology presented by Kasbah et al. (2008) to produce clear specifications, manual refinements, and parallel implementations under VHDL. The developed Datapath presents a rich variety of hardware circuits tailored for achieving specific processing aims and performance characteristics. The developed hardware is tested for accuracy and speed using robust corpuses and compared with software implementations. The main targeted high-performance computing devices are multicore processors for the software version and high-end FPGAs for hardware implementations. The research objectives of this paper are summarized as follows:

1. Exploit the intrinsic parallelism of LB stemmers for Arabic verb root extraction and setting the example for and motivating similar derivations of parallel programs for NLP.
2. Identify and investigate the performance aspects of software implementations of LB stemmers for Arabic verb root extraction while targeting high-performance multi-core processors.
3. Demonstrate a development methodology and investigate the benefits of mapping the derived parallel versions of the adopted algorithm onto hardware. The investigation focuses on the ability of achieving higher throughputs than traditional software implementations by developing multi-cycle and pipelined hardware processors. The hardware developments include the identification of effective implementation specifics and best practices. In addition, the presented research aims at motivating the creation of processors that implement wide NLP features and can be embedded in applications.
4. Present a discussion on the usefulness of the identified software and hardware performance metrics and their usability and applicability in the wider NLP context.
5. Provide enhanced accuracies in analyzing standard Arabic text, such as the *Holy Quran*, by developing algorithms for infix processing.

**Table 2**
Morphological variations of the verb Study (درس) with diacritics showing the active and (passive) voice.

| Subject | Past | Present | Imperative Present | Subjunctive Present | Emphasized Present | Imperative | Emphasized Imperative |
|---|---|---|---|---|---|---|---|
| **I** | دَرَسْتُ<br>(دُرِسْتُ) | أَدْرِسُ<br>(أُدْرَسُ) | أَدْرِسْ<br>(أُدْرَسْ) | أَدْرِسَ<br>(أُدْرَسَ) | أَدْرِسَنّ<br>(أُدْرَسَنّ) | | |
| **We** | دَرَسْنَا<br>(دُرِسْنَا) | نَدْرِسُ<br>(نُدْرَسُ) | نَدْرِسْ<br>(نُدْرَسْ) | نَدْرِسَ<br>(نُدْرَسَ) | نَدْرِسَنّ<br>(نُدْرَسَنّ) | | |
| **You**<br>Male, Singular | دَرَسْتَ<br>(دُرِسْتَ) | تَدْرِسُ<br>(تُدْرَسُ) | تَدْرِسْ<br>(تُدْرَسْ) | تَدْرِسَ<br>(تُدْرَسَ) | تَدْرِسَنّ<br>(تُدْرَسَنّ) | اِدْرِسْ | اِدْرِسَنّ |
| **You**<br>Female, Singular | دَرَسْتِ<br>(دُرِسْتِ) | تَدْرِسِينَ<br>(تُدْرَسِينَ) | تَدْرِسِي<br>(تُدْرَسِي) | تَدْرِسِي<br>(تُدْرَسِي) | تَدْرِسِنّ<br>(تُدْرَسِنّ) | اِدْرِسِي | اِدْرِسِنّ |
| **You**<br>Male, Dual | دَرَسْتُمَا<br>(دُرِسْتُمَا) | تَدْرِسَانِ<br>(تُدْرَسَانِ) | تَدْرِسَا<br>(تُدْرَسَا) | تَدْرِسَا<br>(تُدْرَسَا) | تَدْرِسَان<br>(تُدْرَسَان) | اِدْرِسَا | اِدْرِسَان |
| **You**<br>Female, Dual | دَرَسْتُمَا<br>(دُرِسْتُمَا) | تَدْرِسَانِ<br>(تُدْرَسَانِ) | تَدْرِسَا<br>(تُدْرَسَا) | تَدْرِسَا<br>(تُدْرَسَا) | تَدْرِسَان<br>(تُدْرَسَان) | اِدْرِسَا | اِدْرِسَان |
| **You**<br>Male, Plural | دَرَسْتُم<br>(دُرِسْتُم) | تَدْرِسُونَ<br>(تُدْرَسُونَ) | تَدْرِسُوا<br>(تُدْرَسُوا) | تَدْرِسُوا<br>(تُدْرَسُوا) | تَدْرِسُنّ<br>(تُدْرَسُنّ) | اِدْرِسُوا | اِدْرِسُنّ |
| **You**<br>Female, Plural | دَرَسْتُنّ<br>(دُرِسْتُنّ) | تَدْرِسْنَ<br>(تُدْرَسْنَ) | تَدْرِسْنَ<br>(تُدْرَسْنَ) | تَدْرِسْنَ<br>(تُدْرَسْنَ) | تَدْرِسْنَان<br>(تُدْرَسْنَان) | اِدْرِسْنَ | اِدْرِسْنَان |
| **He** | دَرَسَ<br>(دُرِسَ) | يَدْرِسُ<br>(يُدْرَسُ) | يَدْرِسْ<br>(يُدْرَسْ) | يَدْرِسَ<br>(يُدْرَسَ) | يَدْرِسَنّ<br>(يُدْرَسَنّ) | | |
| **She** | دَرَسَتْ<br>(دُرِسَتْ) | تَدْرِسُ<br>(تُدْرَسُ) | تَدْرِسْ<br>(تُدْرَسْ) | تَدْرِسَ<br>(تُدْرَسَ) | تَدْرِسَنّ<br>(تُدْرَسَنّ) | | |
| **They**<br>Male, Dual | دَرَسَا<br>(دُرِسَا) | يَدْرِسَانِ<br>(يُدْرَسَانِ) | يَدْرِسَا<br>(يُدْرَسَا) | يَدْرِسَا<br>(يُدْرَسَا) | يَدْرِسَان<br>(يُدْرَسَان) | | |
| **They**<br>Female, Dual | دَرَسَتَا<br>(دُرِسَتَا) | تَدْرِسَانِ<br>(تُدْرَسَانِ) | تَدْرِسَا<br>(تُدْرَسَا) | تَدْرِسَا<br>(تُدْرَسَا) | تَدْرِسَان<br>(تُدْرَسَان) | | |
| **They**<br>Male, Plural | دَرَسُوا<br>(دُرِسُوا) | يَدْرِسُونَ<br>(يُدْرَسُونَ) | يَدْرِسُوا<br>(يُدْرَسُوا) | يَدْرِسُوا<br>(يُدْرَسُوا) | يَدْرِسُنّ<br>(يُدْرَسُنّ) | | |
| **They**<br>Female, Plural | دَرَسْنَ<br>(دُرِسْنَ) | يَدْرِسْنَ<br>(يُدْرَسْنَ) | يَدْرِسْنَ<br>(يُدْرَسْنَ) | يَدْرِسْنَ<br>(يُدْرَسْنَ) | يَدْرِسْنَان<br>(يُدْرَسْنَان) | | |

The paper includes thorough performance analysis and evaluation of the developed hardware and compares the findings with similar work in the literature. The evaluation confirms that we have successfully created FPGA cores with accelerated processing throughputs and demonstrated how to take the opportunities to optimize stemming accuracy through infix processing.

This paper is organized so that Section 3 introduces the adopted stemming algorithm for verb extraction, its model of computation, and the concurrent process model. Section 4 presents the processor designs. In Section 5, we present the implementation aspects and challenges. A thorough analysis and evaluation is presented in Section 6 including validation and testing, performance analysis, accuracy analysis, and a general evaluation. Section 7 concludes the paper and sets the ground for future work.

## 3. Unified hardware and software development

The hardware development of the targeted verb root extraction algorithm adopts an informal and systematic approach (Damaj, 2007; Kasbah et al., 2008). The methodology is unified in the sense that it uses common software engineering techniques to model the algorithm; accordingly, software and/or hardware designs are derived and implemented. The steps of development are summarized as follows:

1. Depict the behavior of the model of computation using standard flowcharts.
2. Implement the software version and the hardware using behavioral descriptions.
3. Identify parallel processes and capture the behavior using concurrent process models.
4. Design the Datapath by identifying, allocating, and binding resources.
5. Develop the Finite State Machine (FSM) of the control unit based on the flowcharts.
6. Describe the developed processor using a Hardware Description Language (HDL) and synthesize the implementation for FPGAs.

Although the approach can be used to co-design partitioned versions of the algorithm, the current investigation is intended to produce standalone software implementations and hardware IP-cores for FPGAs.

### 3.1. The model of computation

The targeted algorithm is a standard LB stemmer that extracts the verb root from an input word. In the current investigation, the technical differences between the letters ا and أ are not considered as they do not affect the correctness of the root extractions and the specific purpose of the investigated algorithm. In addition, diacritics are stripped from the input word for the sake of simplicity. The abstract pseudocode of the adopted stemming algorithm is shown in Fig. 1, further expanded into computational steps in Fig. 2, and behavioral software components as shown in Fig. 3. In the developed flowcharts, Java-like descriptions of processes are used.

Generating the stems is a major process in the targeted verb root extraction algorithm. Initially, the *Check Prefixes and Suffixes* and *Produce Pairs* processes produce pairs of all possible combinations of prefixes and suffixes. Then, stem generation takes into consideration the cases of having none, one, or both prefixes and suffixes in the input word from the produced pairs. A list of potential roots is then created by striping all the produced prefix-suffix pairs from the input word. The flowchart of the *Generate Stems* process is shown in Fig. 4. The process *Filter by Size* creates two lists for stems of sizes three (Trilateral) and four (Quadrilateral). Trilateral and quadrilateral stems are the most common among Arabic verbs. Nevertheless, extraction of roots with larger sizes is set as a future work. The last computation step is to compare the generated elements of the two lists of different stem sizes. Finally, the matching root is extracted. For example, processing the word (أفإستسقيناكموها) produces the pairs of all combinations of the prefixes (أفإست) and the suffixes (يناكموها). Many stems are generated based on all possible combinations of prefixes and suffixes that are then striped from the input word and filtered into potential Trilateral and Quadrilateral roots. Among the potential roots produced for (أفإستسقيناكموها) are (تسقي) ,(سقي) ,(سقين), etc. Finally, the root (سقي) is extracted. As for the verb (سيلعبون), the produced potential roots include (لعب) ,(يلعب), and (لعبو); however, the extracted root is (لعب).

### 3.2. Concurrent process model

The development of a concurrent version of the verb extraction algorithm targets parallelism at the process and sub-process levels. At the process level, *Check Prefixes*, *Check Suffixes*, *Produce Prefixes*, and *Produce Suffixes* are scheduled in parallel over two steps (See Fig. 5). In several sub-processes, pleasantly parallel processing is possible. For example, the loop unrolling of the *checkPrefix* process replicates seven parallel comparisons as in Fig. 6; the parallel version of *checkPrefix* returns TRUE if a matching prefix is found. The prefix checks are done in parallel for the first five characters of a 15-character input word (See Fig. 7). The parallel version of the

```
Verb root extraction algorithm
get an input Arabic word
check for all allowable prefixes and suffixes
for all allowable prefixes
        for all allowable suffixes
                produce a (prefix, suffix) pair
for all (prefix, suffix) pair
        generate the stems and filter them by size
for all filtered stems
        compare with stored Arabic verb roots and extract the desired root
```

**Fig. 1.** The verb root extraction algorithm in pseudocode.

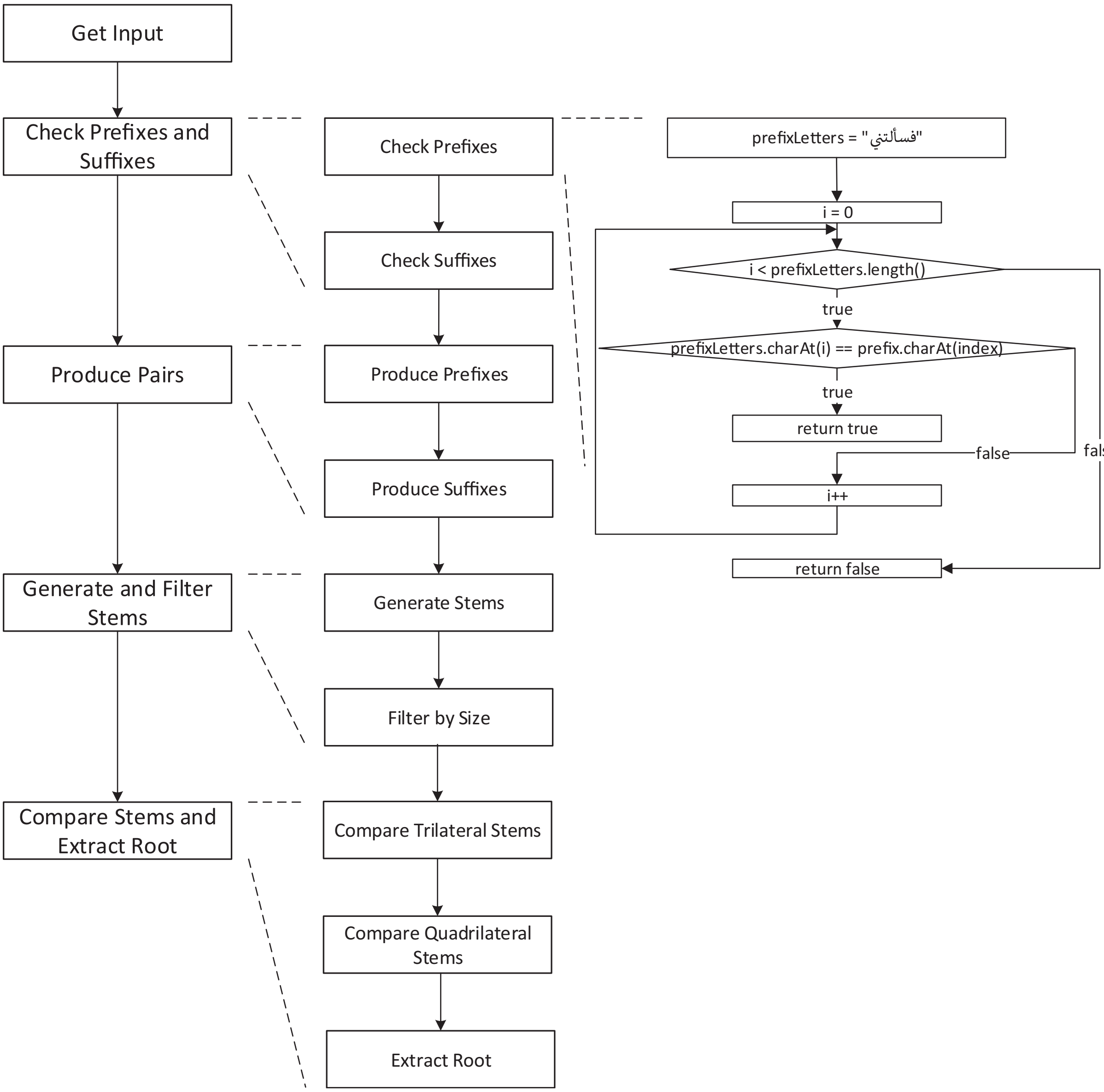

**Fig. 2.** Sample expanded flowcharts of the adopted algorithm.

```
function checkPrefix (char : std_logic_vector(15 downto 0)) return std_logic is
constant prefixes: charsArray(0 to 6) := (x"0623", x"062A", x"0633", x"0641", x"0644",x"0646", x"064A");

    begin
        for i in 0 to 6 loop
            if char = prefixes(i) then
                return '1';
            end if;
        end loop;
```

(a) Behavioral VHDL implementation. The Hexadecimal values of *charsArray* are the Unicode of the Arabic characters ف, س, أ, ل, ت, ن, ي.

```
public static boolean checkPrefix  (String prefix, int index)
    {
    final String prefixLetters = "فسألتني";
        for (int i = 0; i < prefixLetters.length(); i++) {
            if (prefixLetters.charAt(i) == prefix.charAt(index)) {
                return true; }}
        return false;
    }
        return '0';
```

(b) Software implementation under JAVA

**Fig. 3.** Sample implementations of the process *Check Prefixes*.

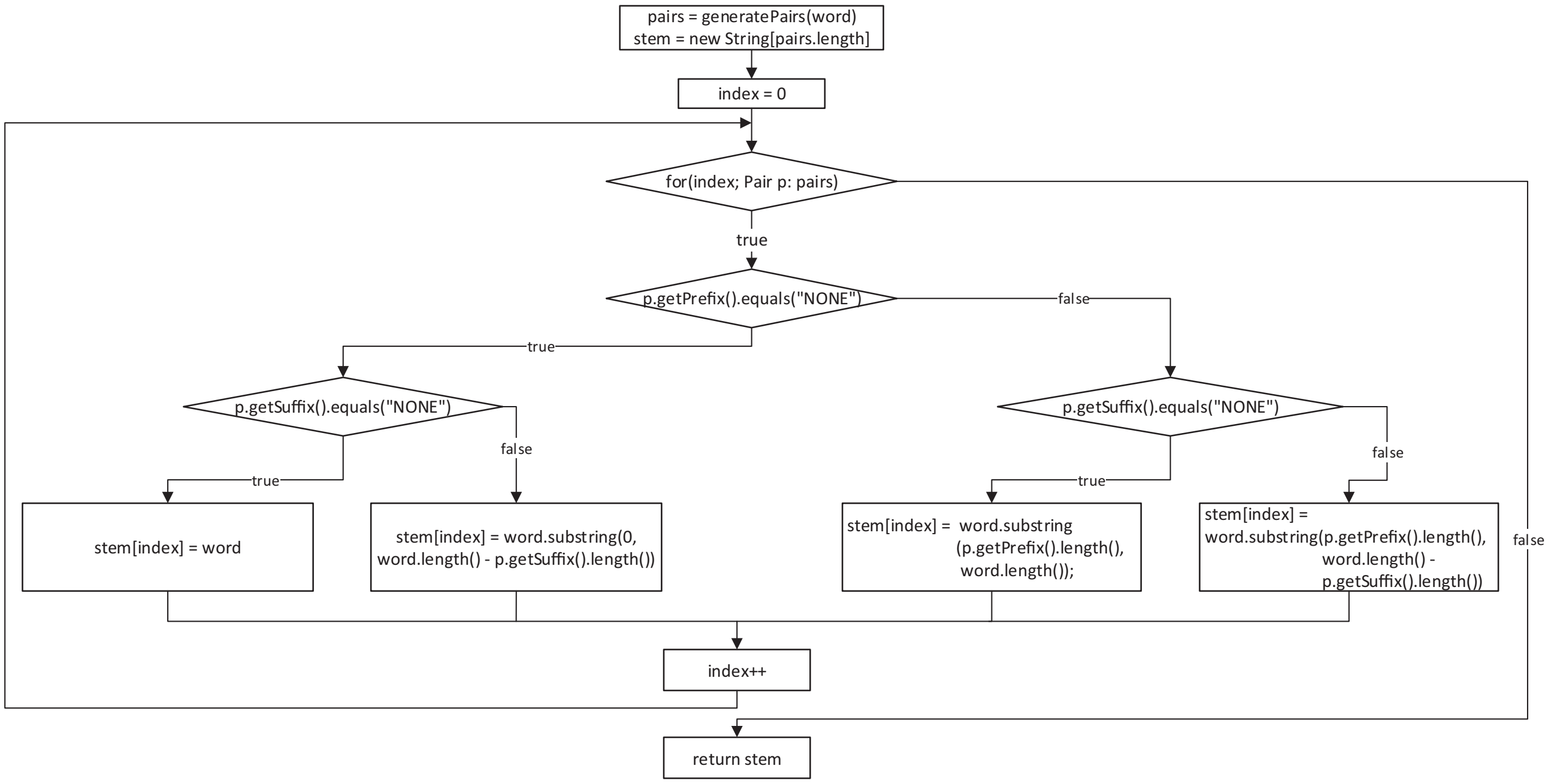

**Fig. 4.** The flowchart of the process *Generate Stems*.

process *Check Suffixes* is developed in a similar fashion to that of *Check Prefixes*; an entity *checkSuffix* is developed to check all the 15-character input word in parallel. The number of input character is chosen based on the longest word in Arabic which is (أفإستسقيناكموها). At this point, prefix and suffix pairs are produced to enable the generation of stems and filtering trilateral and quadrilateral stems by size. Upon the filtering completion, the stems of sizes three and four are compared using parallel instances of *Compare Trilateral Stems* and *Compare Quadrilateral Stems*. However, the compare processes are internally sequential. In addition, the sub-processes of the comparisons are replicated in a data-parallel fashion (See Fig. 8). Finally, the root is extracted. Indeed, more parallelism is possible at the sub-processes level; for example, in *Generate Stems*, the loops can be unrolled into

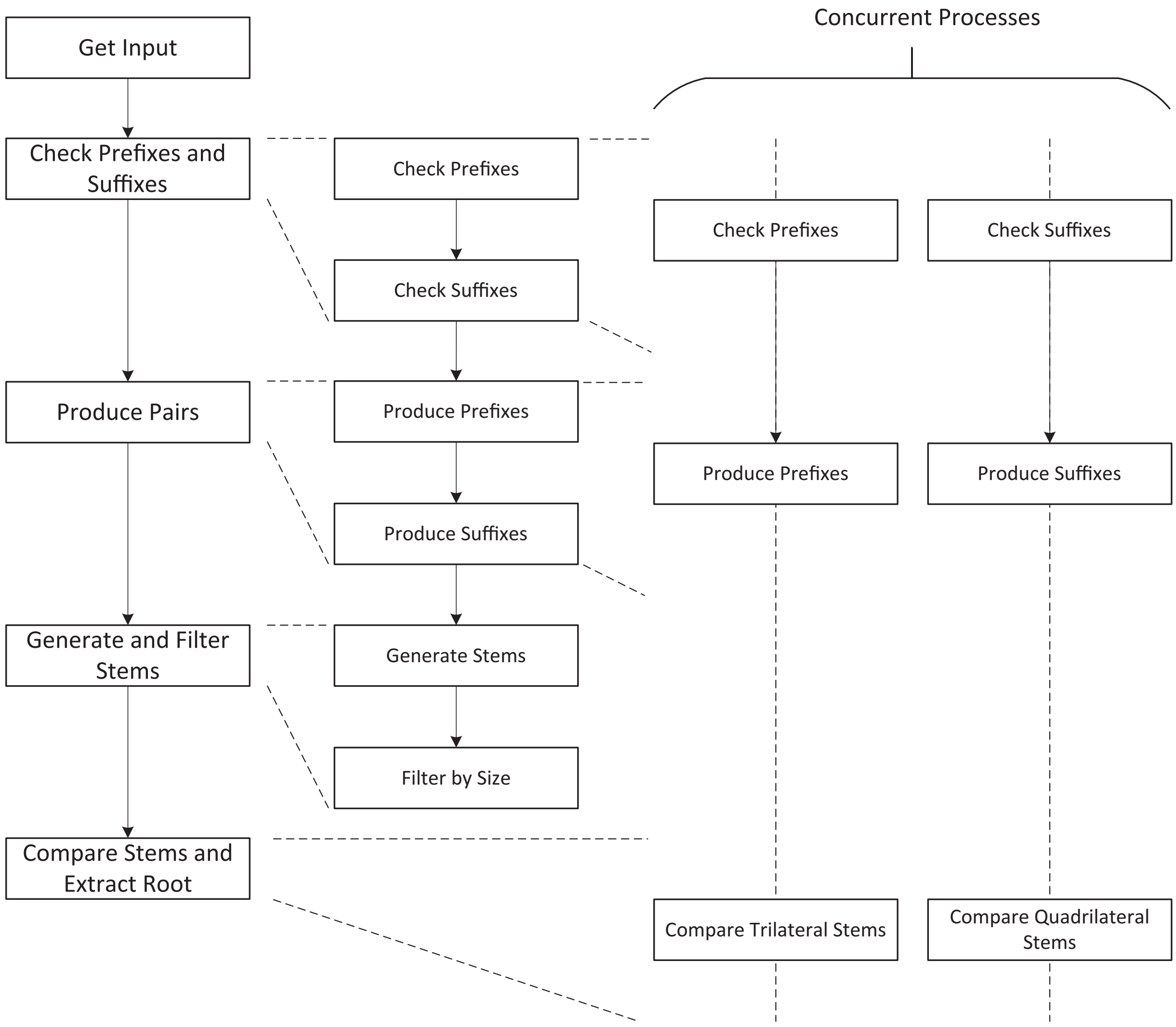

**Fig. 5.** Parallel implementation of the process *Check Prefixes, Check Suffixes, Produce Prefixes, Produce Suffixes, Compare Trilateral Stems*, and *Compare Quadrilateral Stems*.

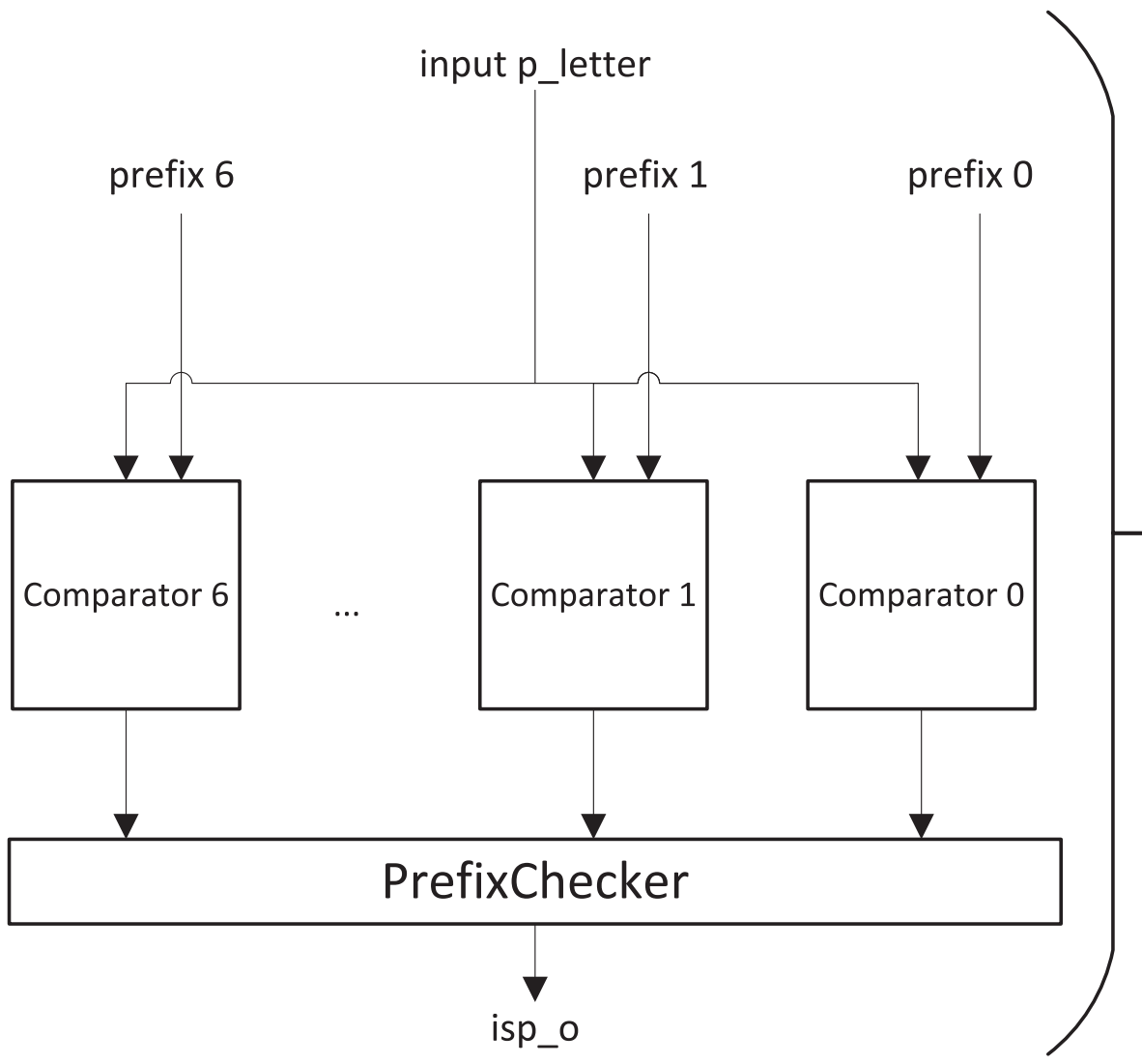

```
entity checkPrefix is
port(         p_letter: in std_logic_vector(15 downto 0);
              isp_o: out std_logic);
end entity;

architecture arch of checkPrefix is
        component comparator_hex is
        port(hex1, hex2: in std_logic_vector(15 downto 0);
              similar: out std_logic);
        end component;

        constant prefixes: charsArray(0 to 6):=
              (x"0623", x"062A", x"0633", x"0641",
               x"0644",x"0646", x"064A");
        signal r : std_logic_vector(6 downto 0);
begin

        Generate_Comparators:for i in 0 to 6 generate
        c_h:    comparator_hex
              port map(p_letter, prefixes(i), r(i));
        end generate Generate_Comparators;

        isp_o <=       r(0) or r(1) or r(2) or r(3) or
                       r(4) or r(5) or r(6);

end architecture;
```

**Fig. 6.** Data parallel implementation of the process *Check Prefixes*. The input to the entity is a potential input prefix character from an input word, while the output is the comparison result.

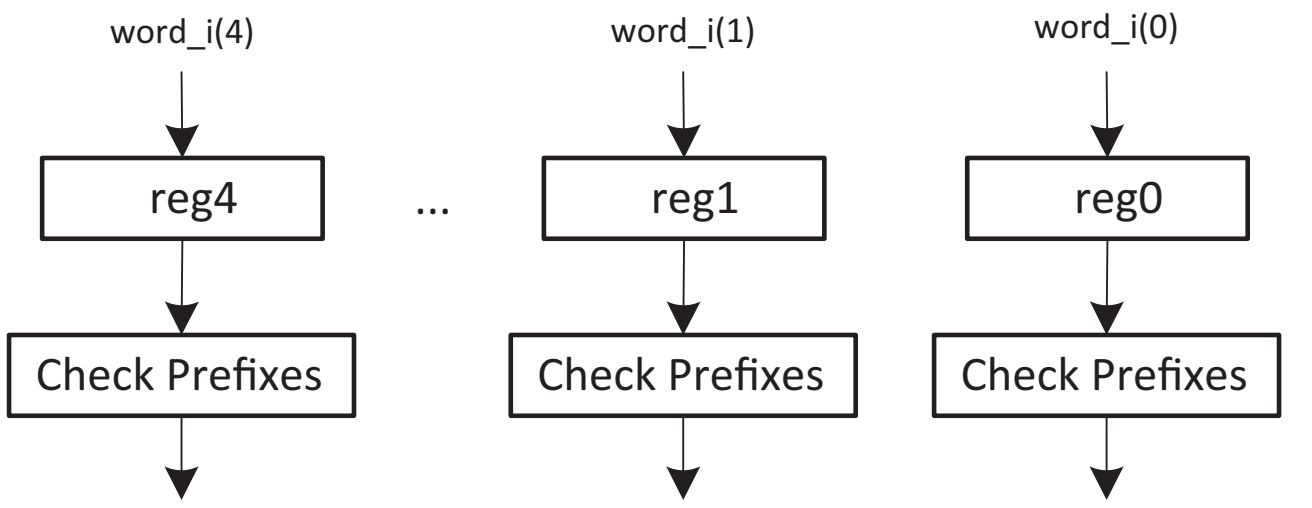

**Fig. 7.** Parallel replication of the entity *checkPrefix*. The first five characters of the input word are initially stored in temporary registers (reg).

different concurrent processes. However, compromising the degree of parallelism is in favor of economizing the use of hardware resources during implementation.

## 4. Processor design

The Arabic verb root extraction hardware is developed as two processors, one multi-cycle and the other is pipelined. The Datapath design comprises identifying, allocating, and binding resources. The Control Unit schedules the execution of resources. Both processors target a total number of five clock cycles to complete their execution.

### 4.1. The Datapath design

The Datapath is built using different types of registers, comparators, and functional units. The design hierarchy is presented in Fig. 9. Different types of registers are used. The purpose of the registers is to store the 16-bit Unicode of a single Arabic character (*regC*), three Arabic characters (*reg3C*), four Arabic characters (*reg4C*), and standard logic general purpose registers of different widths (*reg*). In addition, three different comparators are used for the developed processors including a comparator for single Arabic characters (*comparator*). The unit *stem3_Comparator* compares two Arabic words each of three characters, while the unit *stem4_Comparator* performs the same check but for words of four characters. The functional units *checkPrefix*, *checkSuffix*, *prdPrefixes*, *prdSuffixes*, *generateStems*, and *compareStems* correspond to the processes *Check Prefixes*, *Check Suffixes*, *Produce Prefixes*, *Produce Suffixes*, *Generate and Filter Stems*, and *Compare Stems and Extract Root*. The proposed Datapath is shown in Fig. 10.

The functional units in the Datapath are separated by five arrays of registers. The number of allocated registers is determined per the possible prefixes and suffixes in an Arabic verb. Five registers are allocated for prefixes in the input word, while 15 registers are allocated to examine the suffixes. The examination of the suffixes starts from the left-most character; however, for words shorter than 15, unused (U) character positions are expected. The parallel prefixes and suffixes checks identify the valid characters. The prefix and suffix producers mask any unwanted characters beyond the expected locations. For example, for an input word (يكتبون) the output from the *checkSuffixes* unit is (110111) – with "1" indicating that a suffix is found and a "0" representing that a suffix is not found. At that point, the output is masked to (11UUUU) as the letter (ب), found in the middle of the word, indicates the end of the possibility of having suffixes. Accordingly, all characters before the letter (ب) are masked and output as "U". Producing prefixes works in a similar fashion.

### 4.2. Control unit design

The Control Unit of the verb root extraction processor runs over five states using the five register arrays in the Datapath; this creates independent processing stages that can be pipelined. In this paper, two different control schemes are investigated including pipelined and non-pipelined. The Control Unit Finite State Machine (FSM) of the non-pipelined processor is shown in Fig. 11. The pipelined processor overlaps the execution of all stages. The choice of five stages, and accordingly five clock cycles, is aligned with the number of distinct processing steps and separating register arrays (See Fig. 10).

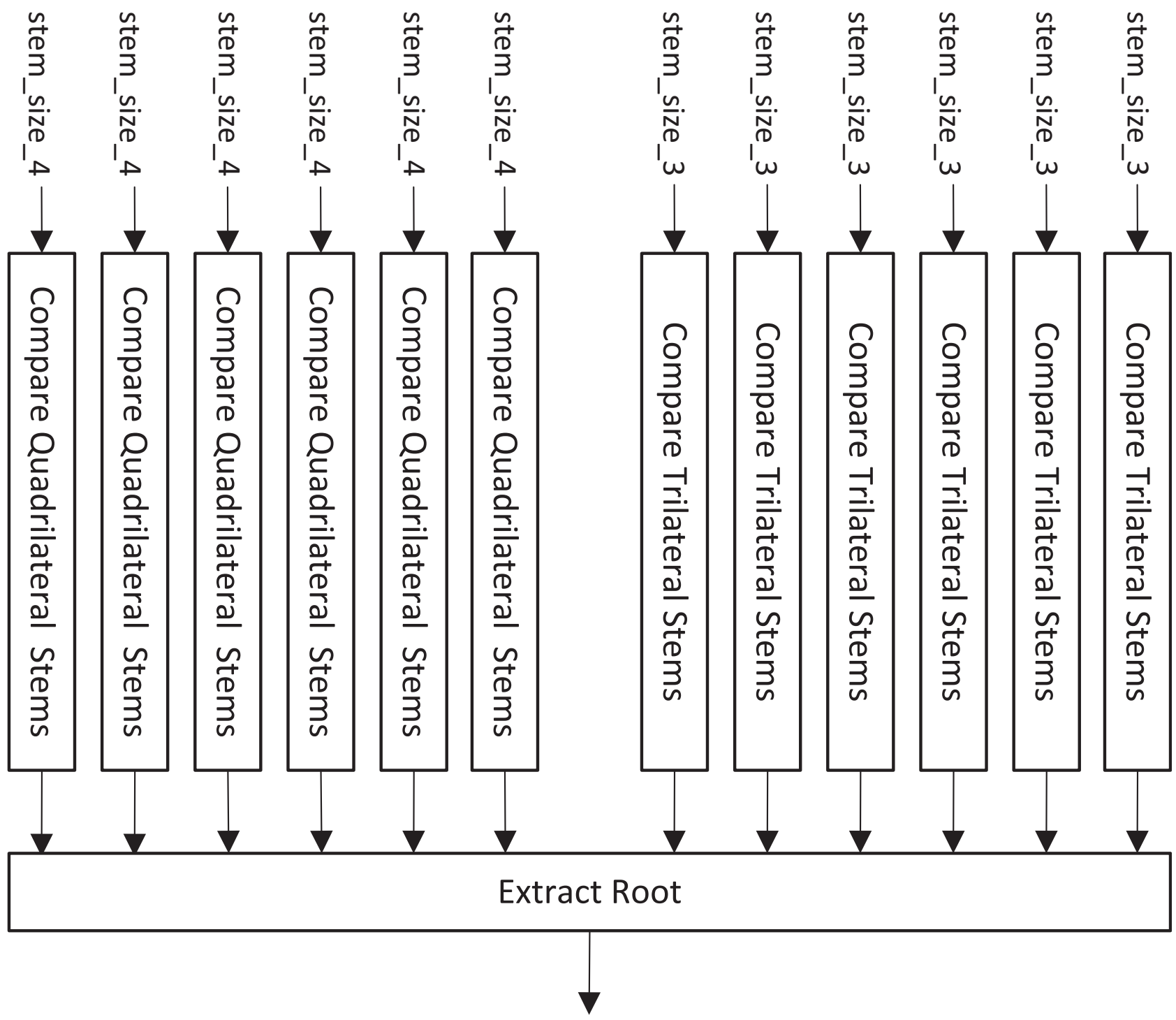

**Fig. 8.** Parallel replication of the processes *Compare Trilateral Stems* and *Compare Quadrilateral Stems*.

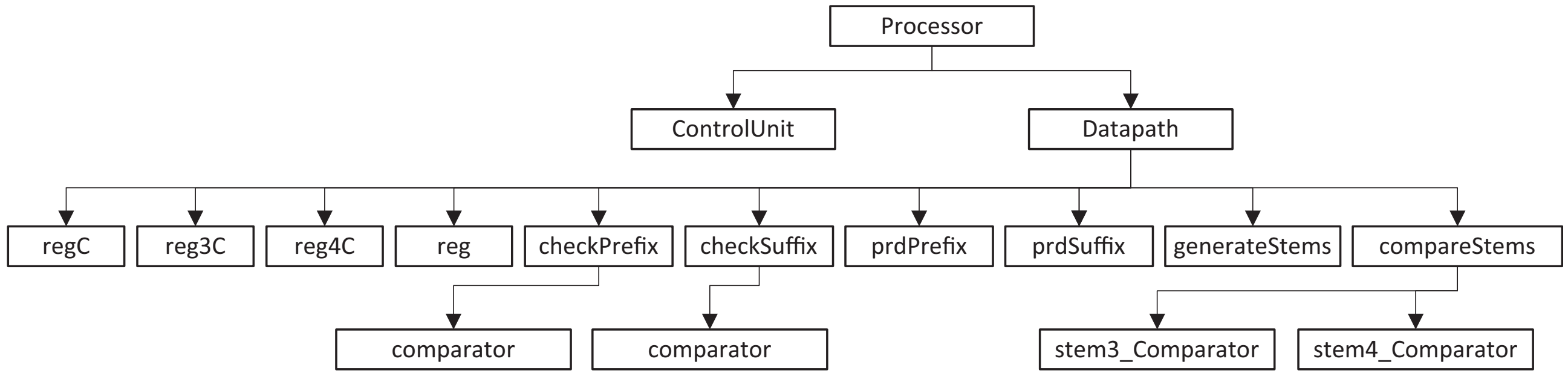

**Fig. 9.** Design hierarchy of the Arabic verb root extraction processor.

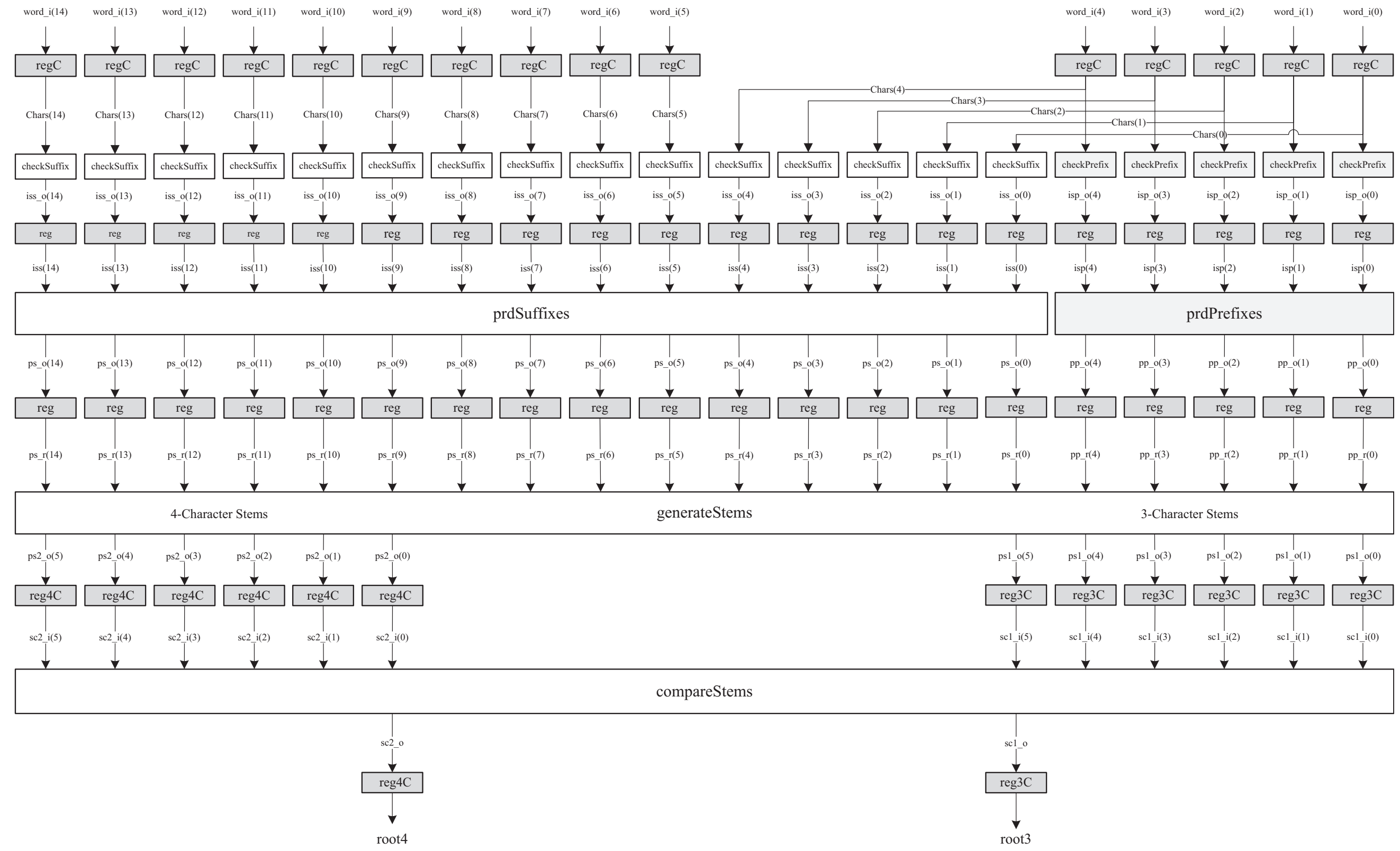

**Signals Legend:**
*word_(i)*: input word;
*iss_o(i)* and *isp_o(i)*: outputs from checkPrefix and check Suffix
*pp_o(i)* and *ps_o(i)*: outputs from prdPrefixes and prdSuffixes
*ps1_o(i)* and *ps2_o(i)*: outputs from generateStems
*sc1_o(i)* and *sc2_o(i)*: outputs from compareStems
*Chars(i)*: input to checkPrefix and checkSuffix;
*iss(i)* and *isp(i)*: inputs to prdPrefixes and prdSuffixes
*pp_r(i)* and *ps_r(i)*: inputs to generateStems
*sc1_i(i)* and *sc2_i(i)*: inputs to compareStems
*root3* and *root4*: output roots

**Fig. 10.** The developed Datapath; showing the binding among different entities using signals. Register arrays are highlighted in dark gray.

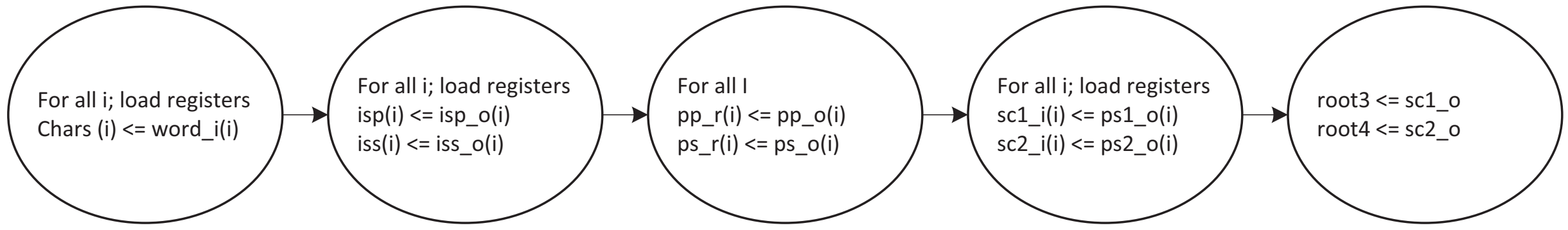

**Fig. 11.** The FSM of the Control Unit.

## 5. Implementation aspects

Several hardware and software implementation aspects are noted for the presented investigation. A variety of implementation tools are used. In addition, specific implementation challenges required the exploration of different problem-solving strategies, such as, substring processing in VHDL and coding of Arabic characters to suite the analysis and implementation options.

### 5.1. Substring truncation in VHDL

The *Generate and Filter Stems* process truncates the input word repeatedly for every produced prefix and suffix pair. The different truncations of the input word are the generated stems that are filtered into two lists per size. Although the filtering is simple, truncation in VHDL requires the development of a special procedure. The proposed truncation procedure examines the output arrays of bits from the *Produce Prefixes* and *Produce Suffixes* processes; the truncation is done at the identified prefix and suffix if the substring included between them is of size 3 or 4 characters. The main constructs in the proposed procedure are two loops to check for all possible prefix and suffix pairs, a substring size equation, and an assignment of the truncated substring. The code segment in VHDL is shown in Fig. 12. The size of the substring to be truncated is calculated as the difference between the prefix and suffix indices. The prefix index can vary between 0 and 5, while the suffix index can vary between 15 down to 0. Table 3 shows all the permitted substrings of the word (سيلعبون) with the produced lists of prefixes as (0000011) and suffixes as (110000).

The pleasant data parallelization of the *Substring Truncation* code segment in Fig. 12 is possible; however mass replications are expected. The replication should consider all possible combinations of prefix-suffix pairs. The pleasantly parallel version will be able to generate and filter all stems in a few number of steps, but at the cost of a large hardware circuit size. Indeed, the parallelization of all the presented processes can lead to interesting improvements in the processing performance.

### 5.2. Coding of Arabic characters

Two different codes for Arabic characters are used in the presented hardware and software implementations. Arabic Unicode is used for processing in both hardware and software implementations; however, an ASCII-based code is created to aid displaying the tested words in the simulation tools. For example, the character (س) is processed in its Unicode ($0633_{hex}$) and displayed as (Sin) in the simulator. The VHDL implementation of the Arabic Unicode is realized as a new type in Package with the following declaration of a list of 16 bits under the user-defined type *charsArray*:

```
constant p_index : intArray(0 to 5):= (-1, 0, 1, 2, 3, 4);
constant s_index: intArray(15 downto 0):=(15, 14, 13, 12, 11, 10, 9, 8, 7, 6, 5, 4, 3, 2, 1,0);

begin

process (word_i, pp_r, ps_r)
variable count1, count2: integer range 0 to 5 := 0;
variable found1, found2: std_logic:= '0';

begin

count1 := 0;
count2 := 0;
for i in 0 to 5 loop
        found1 := '0';
        found2 := '0';
        if pp_r(i) = '1' then
                for j in 15 downto 0 loop
                        if ps_r(j) ='1' then
                                if (s_index(j) - 1) - (p_index(i) + 1) = 2 then
                                        stem3_o(count1) <= word_i( (p_index(i) + 1) to (s_index(j) - 1)
                                        if( count1 < 5) then
                                                count1 := count1 + 1;
                                        end if;
                                        found1 := '1';
                                elsif ((s_index(j) - 1) - (p_index(i) + 1)) = 3 then
                                        stem4_o(count2) <= word_i( (p_index(i) + 1) to (s_index(j) - 1) );
end process;
end architecture;
```

**Fig. 12.** The substring truncation and filtering of the process *Generate Stems* in VHDL.

**Table 3**
Truncation of stem substrings of the verb (سيلعبون).

| Input Word | ن | و | ب | ع | ل | ي | س | s_index | p_index | (s_index(j) – 1) – (p_index(i) + 1) |
|---|---|---|---|---|---|---|---|---|---|---|
| *Produce Prefixes* Output | 0 | 0 | 0 | 0 | 0 | 1 | 1 | | | |
| *Produce Suffixes* Output | 1 | 1 | 0 | 0 | 0 | 0 | 0 | | | |
| 1. Trilateral Stem 1 | | | ب | ع | ل | | | 5 | 1 | 2 |
| 2. Quadrilateral Stem 1 | | | ب | ع | ل | ي | | 5 | 0 | 3 |
| 3. Quadrilateral Stem 2 | | و | ب | ع | ل | | | 6 | 2 | 3 |

*type charsArray is array (natural range<>) of std_logic_vector(15 downto 0)*;

### 5.3. Implementation and analysis tools

A variety of tools are used to implement, test, and analyze the developed hardware and software implementations. The investigation targets two high performance computing systems, namely, the *Dell Precision T7600* with its six-core *Xeon* processor and 32 GB of RAM, and *Altera STRATIX-IV* FPGA. The tools used for hardware analysis are *Quartus* and *ModelSim-Altera*, and *Intel VTune Amplifier* for software analysis. The hardware implementations are done using VHDL. The software version is implemented using Java under *Netbeans*.

## 6. Analysis and evaluation

### 6.1. Validation and testing

The presented hardware and software implementations are validated through intensive testing at the component, integration, and system levels. Sample system executions of the developed non-pipelined processor under *ModelSim* are shown in Figs. 13 and 14. In Fig. 13, the successful root extraction of the complicated Arabic verb (أفإستسقيناكموها) is shown; the extracted trilateral root is (سقي). Fig. 14 shows the extraction for the input Arabic verb (فتزحزحت); the extracted quadrilateral root is (زحزح). Sample output for the pipelined processor is shown in Fig. 15; the extracted roots appear after the fifth cycle and then every cycle. Indeed, the adopted language rules and roots are verified by language specialists.

The testing of the developed implementations is extended to include two formal linguistic reference corpora that comprise the text of the complete *Holy Quran* and an individual test of its 29th Chapter, namely *Surat Al-Ankabut* (The Spider Chapter). The text of the *Holy Quran* includes 77476 words or 17622 words without repetition. The total number of roots that can be extracted from Quran is 1767. *Surat Al-Ankabut* has 980 words Khodor and Zaki (2011). Slightly different word counts of the *Holy Quran* are reported in the literature due to the difference of classification and definition of what is a word.

### 6.2. Performance analysis

A thorough performance analysis is performed for the software and hardware implementations. The software implementation is evaluated per the following metrics (Damaj and Kasbah, 2017):

- Execution Time (ET): The time between the start and the completion of execution.
- Throughput (TH): The total amount of work done per time; in the current investigation, it is defined as the number of processed Words per Seconds (Wps).

The hardware implementation is evaluated per ET, TH, and the following additional metrics (Damaj and Kasbah, 2017):

- Propagation Delay (PD): The time required for a signal from an input pin to propagate through combinational logic and appear at an external output pin.
- Look-Up Table (LUT): The number of combinational adaptive lookup tables required to implement an algorithm in hardware. The number of LUTs is an indicator of the size of hardware in

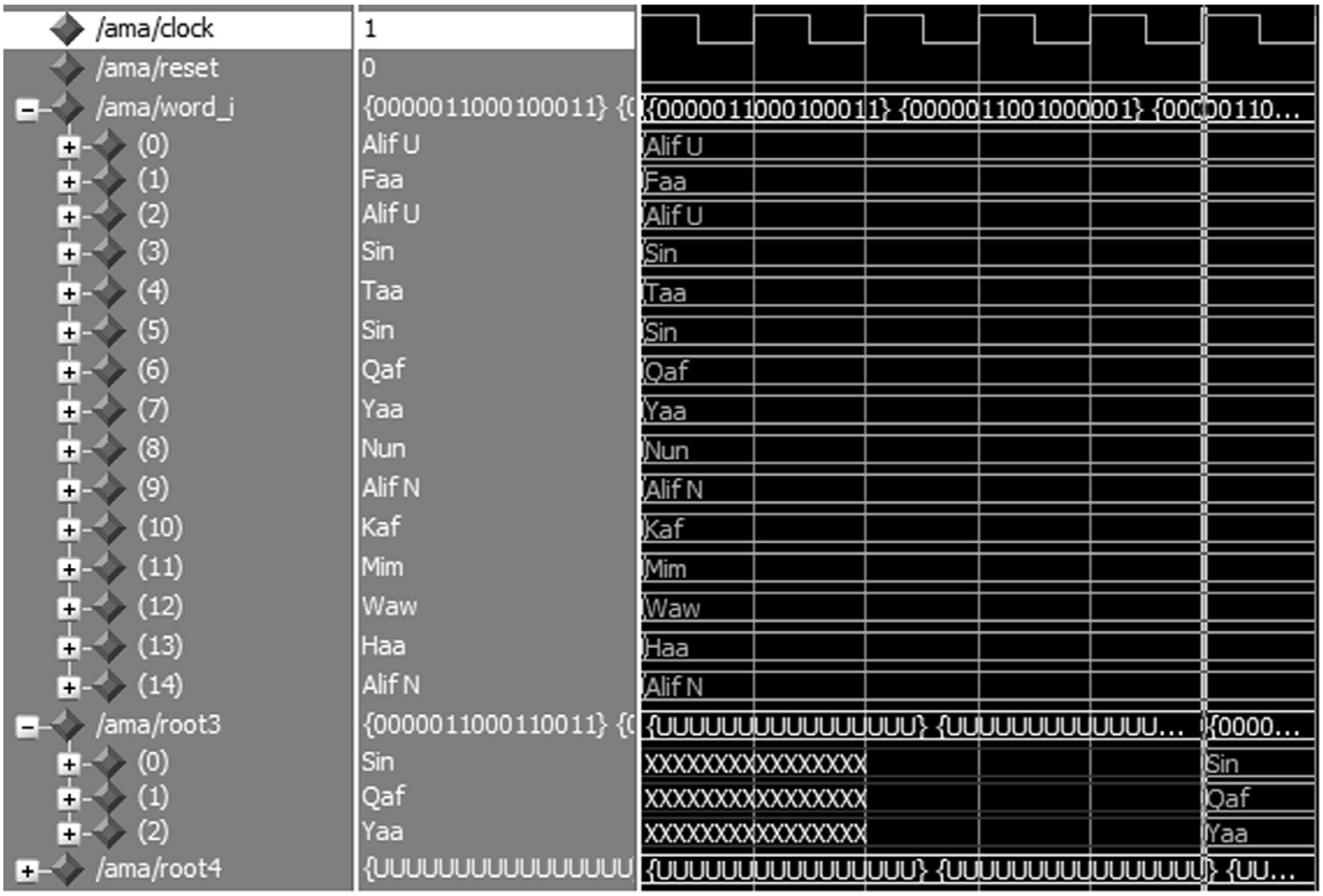

Fig. 13. ModelSim output of the root extraction of the verb (فأستسقيناكموها) and the output root is (سقي).

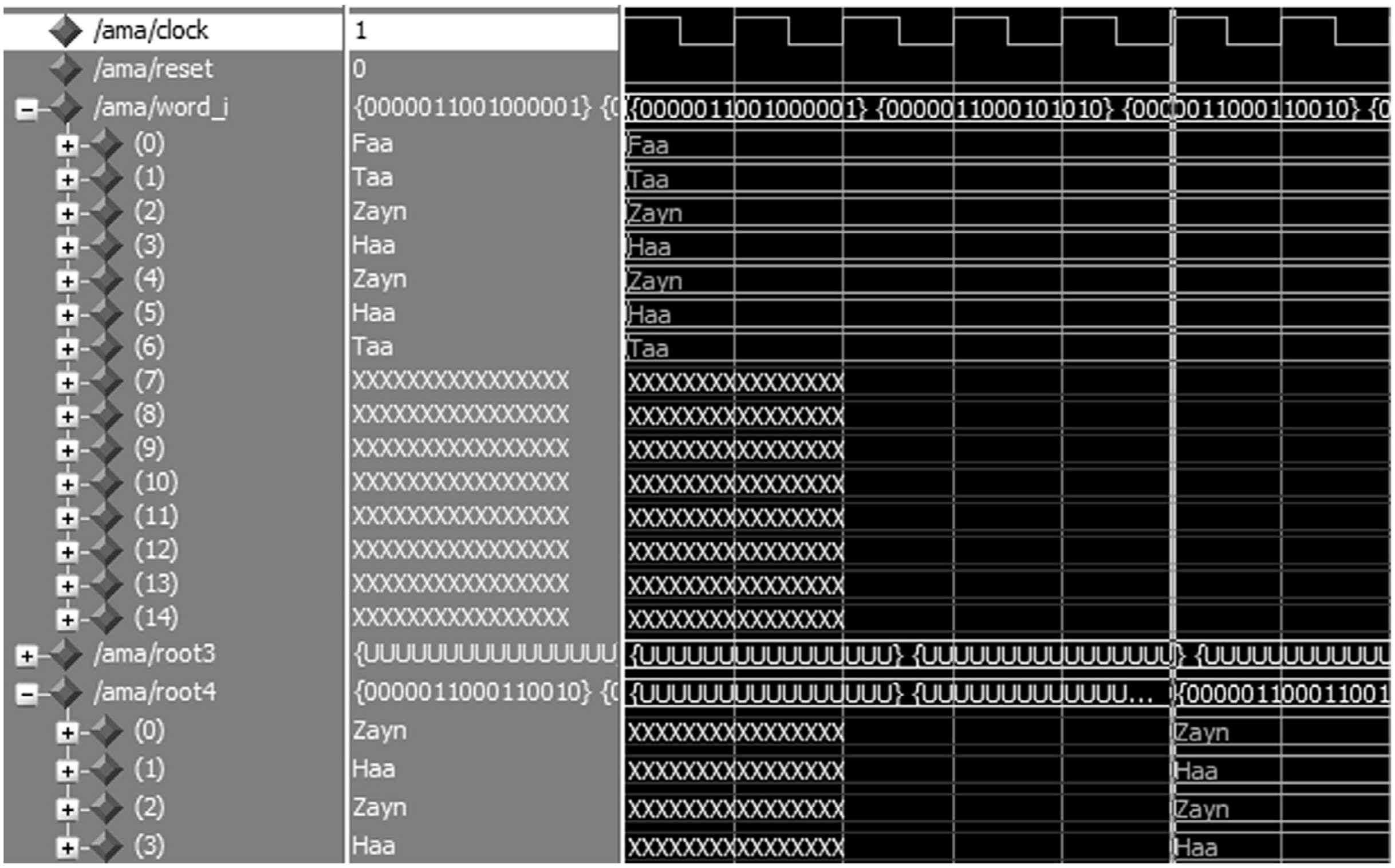

**Fig. 14.** ModelSim output of the root extraction of the verb (فتزحزحت) and the output root is (زحزح).

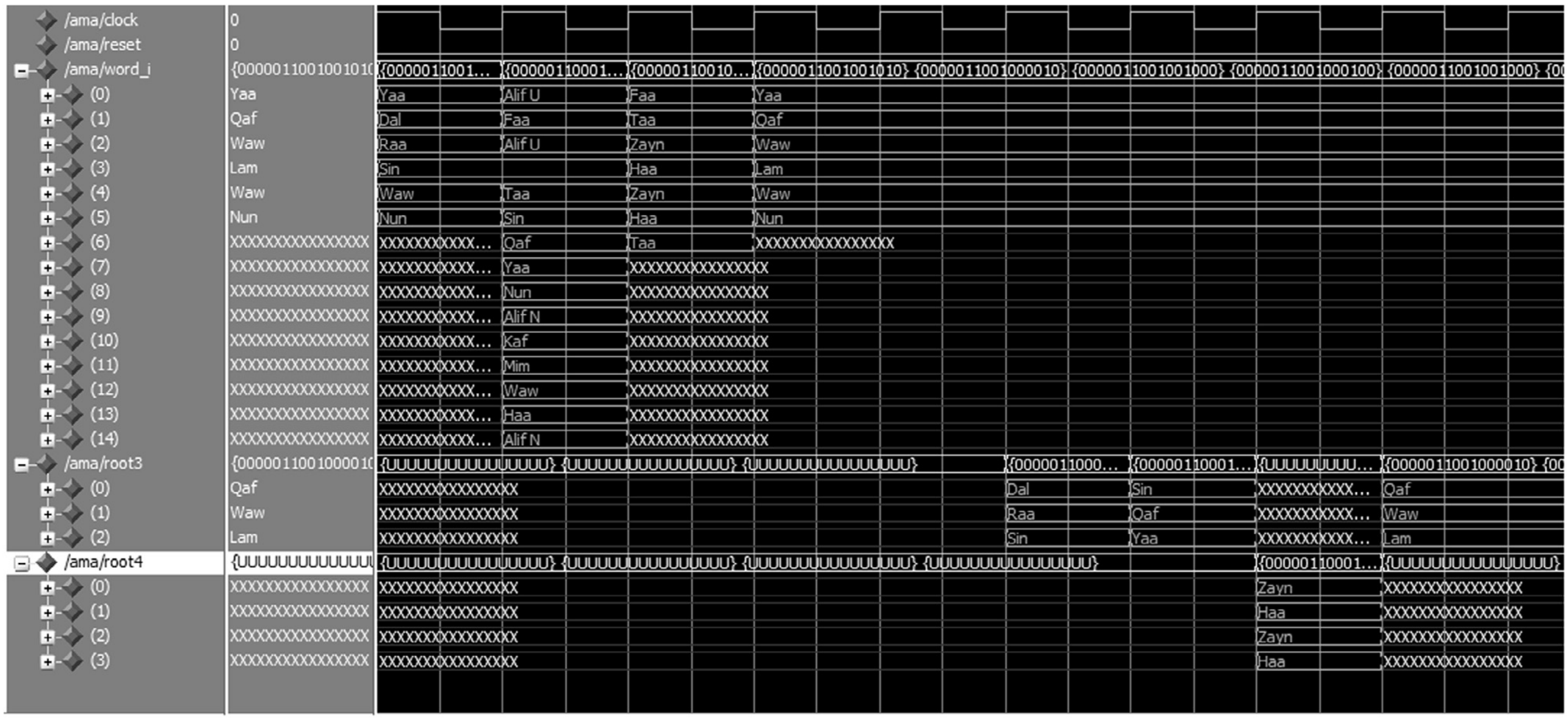

**Fig. 15.** ModelSim output of the root extraction of several verbs using the pipelined processor.

Altera devices. In other devices, the area could be measured in terms the total number of gates, logic elements, slices, etc.

- Logic Register (LR): the total number of logic registers in the design.
- Power Consumption (PC): The power consumption of the developed hardware in Watts.

The software implementation achieved a maximum Throughput of 373.3 Wps. The non-pipelined processor achieved a Throughput of 2.08 MWps. With respect to the ratio of Throughputs, the non-pipelined processor achieved a speedup of 5571 times higher than the software implementation. However, the pipelined processor achieved a speedup of 28873.51 times higher than the software implementation. Moreover, the hardware implementations enjoyed a fixed 5-cycle implementation but a somewhat low maximum frequency of 10.4 MHz for the non-pipelined processor and 10.78 MHz for the pipelined processor. The reported frequencies were limited due to hold checks in the synthesized circuit. Tables 4 and 5 summarize the hardware analysis results for both processors, while Fig. 16 diagrams the throughputs.

The pipelined implementation can lead to better performance for extended number of input words. Fig. 17 plots the projected throughput speedups of the pipelined processor over the non-pipelined processor with respect to the change of the number of analyzed input words. The pipelined processor achieved 10.78 MWps and 10.73 MWps for processing the texts of the *Holy Quran* and *Surat Al-Ankabut*; the throughputs achieved speedups of 5.18 and 5.16 over the non-pipelined implementation and speedups of 28873.5 and 28757.6 over the software implementation.

### 6.3. Infixes processing and accuracy analysis

The accuracy of LB stemmers can be significantly improved with considering the infixes in the Arabic verb root extraction. The way we incorporate the processing of infixes is by considering the most common cases. For instance, in Khodor and Zaki (2011) the authors classified the verb root Say (قول) as one of the most repeated roots in the *Holy Quran* with a frequency of 1722 occurrences. In addition, the derived morph "Then they said" (فقالوا) is considered as one of the most repeated words with 255 occurrences in the *Holy Quran*. The proposed incorporation of infix processing is based on frequency analysis of Arabic verbs as they appear in formal texts like the *Holy Quran*.

Two algorithms are developed to reduce or replace infixes in the extracted roots from LB stemming; the proposed algorithms led to significant improvement in the successful extraction of the verb roots. The algorithms are incorporated as processes running after the lists of Trilateral and Quadrilaterals are filtered, compared, and the root is not found. The process *Remove Infix* checks the second character of trilateral and quadrilateral stems, using a process *Check Infixes,* to possibly extract bilateral and trilateral roots. The algorithm of *Remove Infix* is defined in pseudocode as shown in Fig. 18. Example extractions are of the bilateral verb (حج) from the trilateral verb (حاج) and the trilateral verb root Wrote (كتب) from the quadrilateral stem Corresponded With (كاتب).

The process *Restore Original Form* handles the common case of converting the infix of a verb root from (و) to (ا). The developed process restores the original form by reversing the conversion. Example conversion is for the highly frequent root (قول) from the variation (قال). The pseudocode description of the process *Restore to Original Form* is shown in Fig. 19. Indeed, the processing of infixes can be addressed in a larger scale to further increase the accuracy or automatic verb root extraction in Arabic MA.

The accuracy of analyzing the *Holy Quran* text using the software implementation with and without infix processing is shown in Table 6. The number of successful roots extracted with the infix processing procedure increased the accuracy to 87.7% from an initial measurement of 71.3%. Table 7 presents the top frequencies of verbs extracted from the Holy Quran text in comparison to Khoja stemmer (Khoja, 2017; Khoja and Garside, 1999). In many instances, Khoja stemmer achieved higher accuracy but not for the root (كون) where the proposed implementation attained a 53% higher accuracy. The analysis of *Surat Al-Ankabut* text attained an accuracy of 90.7%. The accuracies reported for analyzing the nouns and verbs of *Surat Al-Ankabut* in Sawalha and Atwell (2008) are 62.27% for Khoja Stemmer (Khoja, 2017; Khoja and Garside, 1999), 57.16% for Tim Buckwalter Arabic Morphological Analyzer (Buckwalter, 2002), and 58.7% for the Voting Algorithm presented in Sawalha and Atwell (2008).

**Table 4**
Hardware analysis results under STRATIX IV FPGA.

| Metrics | Non-Pipelined Processor | Pipelined Processor |
|---|---|---|
| Fmax (MHz) | 10.4 | 10.78 |
| LUT | 85895 (47% Utilization) | 70985 (39% Utilization) |
| LR | 853 (<1% Utilization) | 1057 (<1% Utilization) |
| Power Consumption (mW) | 1006.26 | 1010.96 |

**Table 5**
The Throughput to hardware area ratios of the non-pipelined and the pipelined processors.

| Metrics | Non-Pipelined Processor | Pipelined Processor |
|---|---|---|
| The text of the Holy Quran | | |
| Throughput to LUT Ratio (Wps/ALUTs) | 24.22 | 151.85 |
| Throughput to LR Ratio (Wps/LR) | 2438 | 10197 |
| The text of Surat Al-Ankabut | | |
| Throughput to LUT Ratio (Wps/ALUTs) | 24.21 | 150.6 |
| Throughput to LR Ratio (Wps/LR) | 1967.83 | 10116.09 |

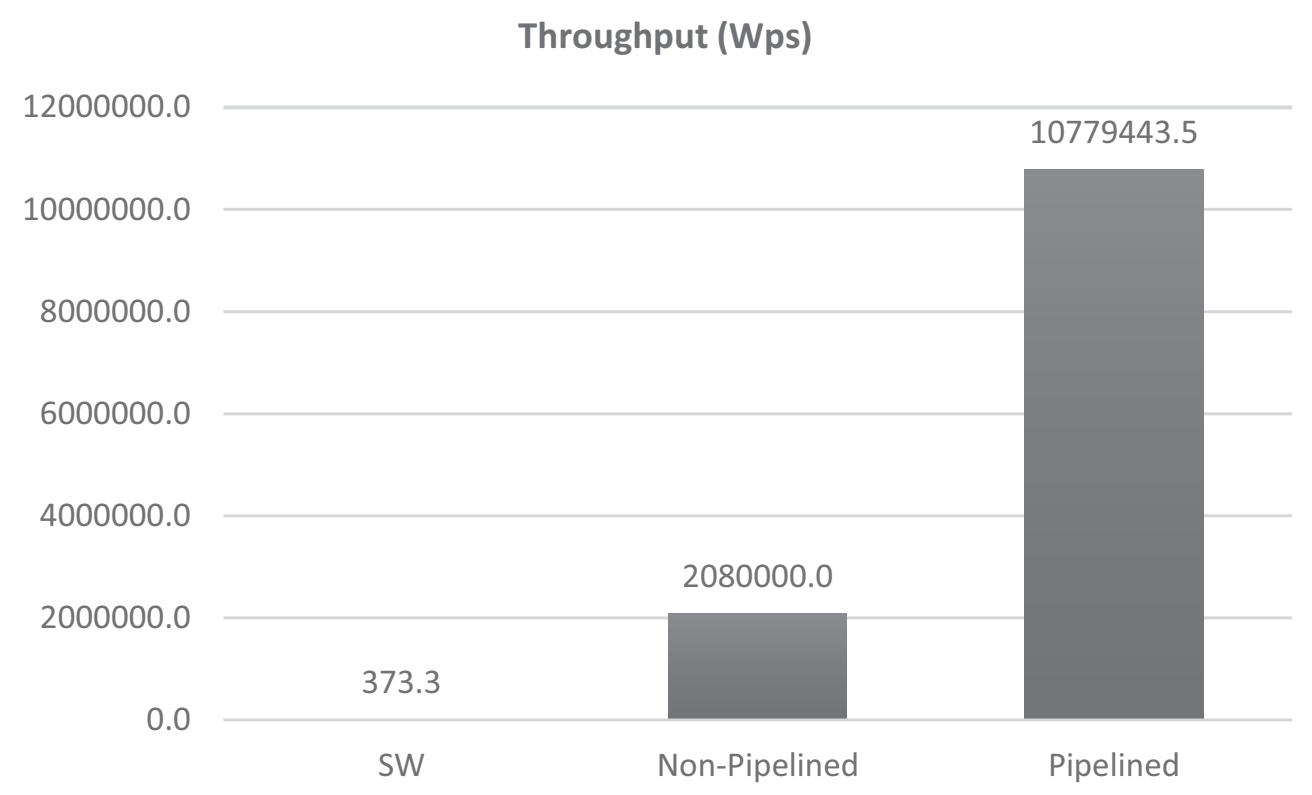

**Fig. 16.** Throughput of the different system implementations of the analysis of the *Holy Quran* text.

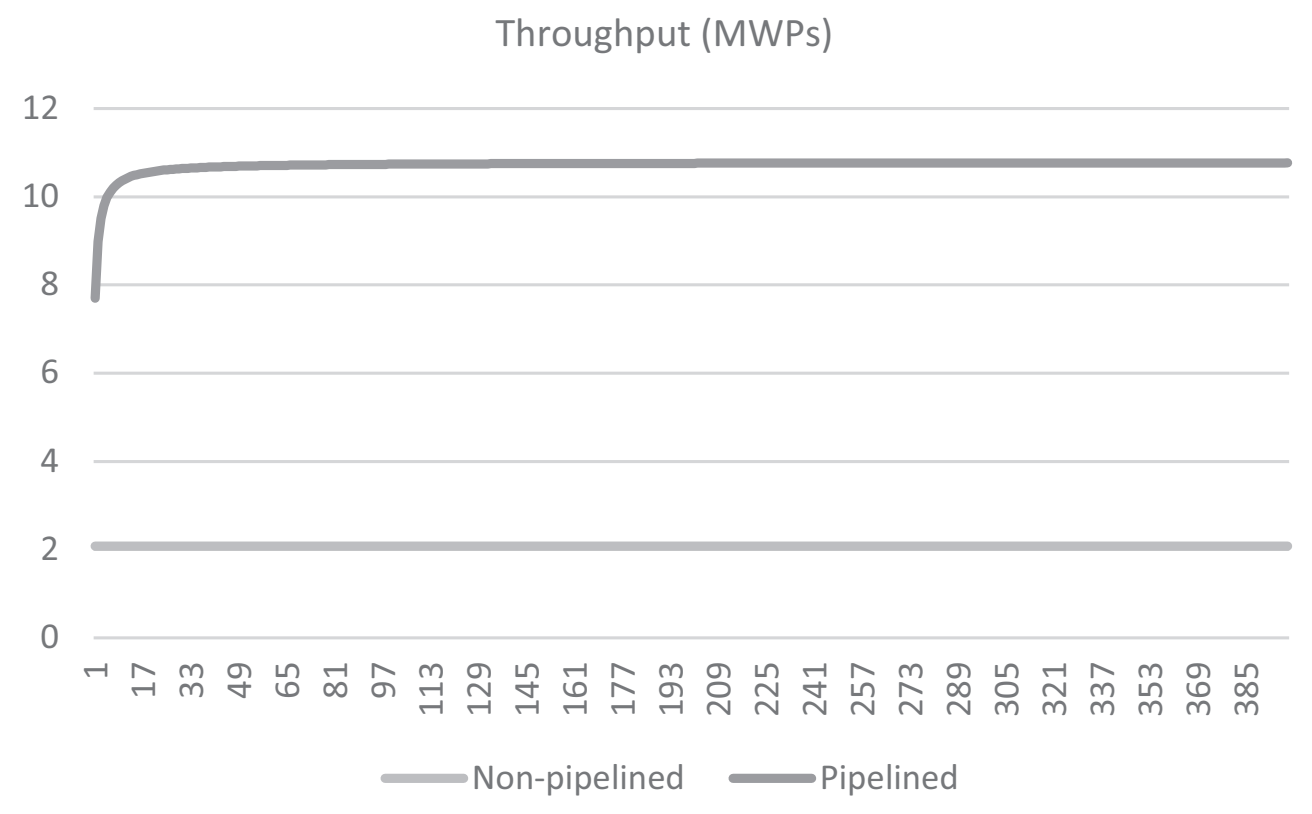

**Fig. 17.** The Throughput change wrt the change in number of input words for the non-pipelined and pipelined processors.

### 6.4. General evaluation

The intrinsic concurrency and the computational complexity of Arabic language have greatly inspired the current investigation of parallel hardware development for faster morphological analysis. The current investigation is pioneering in the fact that limited work has been reported in the literature aiming at creating parallel hardware cores suitable for FPGA implementations and for Arabic

***Remove infix algorithm***
*for all trilateral and quadrilateral stems*
*if the second character is an infix*
*remove character from stem*
*compare the reduced stems and extract root*

**Fig. 18.** The Remove Infix Algorithm.

***Restore original form algorithm***
*for all trilateral stems*
*if the second character is (ا)*
*replace it with (و)*
*compare the stems and extract root*

**Fig. 19.** The Remove Original Form Algorithm.

**Table 6**
The analysis results of the *Holy Quran* text using the software implementation.

| Analysis of the Holy Quran Text | No. of Extracted Verb Roots | Accuracy (%) |
|---|---|---|
| Without Infix Processing | 1261 | 71.3% |
| With Infix Processing | 1549 | 87.7% |

language. In the literature, application-specific hardware systems are mainly proposed to implement high-speed string matching algorithms (Murty et al., 2003; Grossi, 1992; Ratha et al., 2000; Raman and Shaji, 1995). With no doubt, the developed hardware cores for Arabic language can be refined to suite Application-specific Integrated Circuits (ASICs). The development of the hardware and software implementations adopted an easy-to-use hardware/software co-design methodology that employs flowcharts, concurrent process models, and Datapath and FSM diagrams to develop the various implementations. The developed models enabled reasoning and straightforward parallelization of the algorithm.

The developed LB stemming algorithm has several advantages and opportunities for improvement. The developed algorithm has a quadratic complexity $O(n^2)$. However, the complexity is reduced to a constant $(c)$ growth with the change in input size $O(c)$ for different sub-processes in the parallel version. Still, it is expected that the proposed parallel design undertakes a quadratic growth due to the process *Compare Stems and Extract Root*. The process can be reduced to a logarithmic complexity $O(log(n))$ if a tree-based search is used. Several out-of-the-box algorithms are developed to match the best possible degree of parallelism in the concurrent model and respond to the specifics of the developed system. For example, the Substring Truncation algorithm of Section 4.1 had to deal with the fact that we are dealing with two parallel registers to identify infixes and suffixes and accordingly cut the input word. Indeed, LB stemmers obviously rely on the information found in the letters of the input word. Such a reliance on the input word can lead to major complexities and inaccuracies to deal with roots that are often transformed by replacement, fusion, inversion, or deletion (Yagi and Harous, 2003).

The software implementation is simple, scalable, and can be easily upgraded with additional root-extraction rules. The software implementation with infix processing achieved high accuracies in analyzing the texts of the targeted corpora. The difference in erroneous extractions was small in comparison to Khoja Stemmer in analyzing several high-frequency Arabic verbs in the targeted corpuses. In certain cases, the proposed algorithm outperformed Khoja Stemmer in accuracy.

The hardware implementations enable the creation of non-pipelined and pipelined processors. The developed processors enjoy the parallel models and enable high throughputs when mapped onto FPGAs. Although the hardware cores can operate on relatively low frequencies, around 10.5 MHz, the throughput-to-logic-area ratios are relatively high; this indicates high-speeds and adequate logic area use. The targeting of hardware cores with higher throughputs is challenged by the sequential processing within specific processes. In addition, to achieve higher throughputs the restriction to five clock cycles should be compromised. The processes *Generate Stems*, *Filter by Size*, and *Compare Stems* can be redeveloped for higher degrees of parallelism; however, such a redevelopment will require additional hardware resources. Indeed, smaller propagation delays and accordingly higher clock rates are expected with the increase in degree of parallelism and the break of the critical path into cycle counts that are higher than five.

**Table 7**
The analysis results of the *Holy Quran* text using the software implementation.

| Root | Actual | Khoja (1) | Proposed Alg. with Infix Proc. (2) | Absolute% Difference between (1) and (2) | Proposed Alg. without Infix Proc. |
|---|---|---|---|---|---|
| علم | 854 | 798 | 592 | 24% | 436 |
| كفر | 525 | 521 | 376 | 28% | 300 |
| قول | 1722 | 1195 | 1022 | 10% | 267 |
| نفس | 298 | 289 | 256 | 11% | 254 |
| نزل | 293 | 274 | 230 | 15% | 230 |
| عمل | 360 | 356 | 273 | 23% | 225 |
| خلق | 261 | 251 | 216 | 13% | 206 |
| جعل | 346 | 307 | 207 | 29% | 203 |
| كذب | 282 | 262 | 214 | 17% | 190 |
| كون | 1390 | 32 | 765 | 53% | 161 |

This paper calls for broadening the theoretical NLP discussions to include parallel processing and hardware/software co-design. In practice, hardware acceleration of NLP algorithms is currently facilitated with the availability of high-performance multi-core processors and a variety of high-end co-processing options, such as, FPGAs, Graphical Processing Units (GPUs), etc. The proposed parallelization of the LB stemmer, for Arabic verb root extraction, confirms its usefulness in implementing high-speed hardware implementations. The identified metrics, namely, degree of parallelism, TH, ET, LUT, LR, and PC create additional practical and effective performance profile for analysis and evaluation of NLP algorithms and implementations. Furthermore, the proposed infix processing algorithms provide enhanced accuracies in analyzing standard Arabic text, such as, the *Holy Quran*. With no doubt, the adopted methodology, development patterns, implementation specifics, enhanced algorithms, identified metrics, etc. are applicable in the wider NLP context and sets the ground for a wider incorporation of parallel processing and development of extended NLP processors.

## 7. Conclusions

Morphological analysis in Arabic language is computationally intensive, has numerous forms and rules, and intrinsically parallel. The aim of the presented work is successfully achieved by applying an effective modelling technique, developing parallel processes, and deriving pioneering implementations with appealing performance characteristics. The developed parallel processor attained a speed of 2.08 MWps with a speedup of 5571.4 times higher than the software implementation. The developed pipelined processor achieved a speedup of 5.18 times higher than the non-pipelined core. The created hardware cores run at frequencies of 10.4 MHz (non-pipelined) and 10.78 MHz (pipelined), and achieved throughput-to-area ratios of 151.85 Wps/ALUT and 10197 Wps/LR for the pipelined core. The proposed algorithm featured enhanced analysis options over traditional LB stemmers to process infixes with accuracies of 87% and 90.7% for analyzing the texts of the *Holy Quran* and its Chapter 29 – *Surat Al-Ankabut*. Future work includes developing concurrent models with increased degree of parallelism. Future work also includes the optimization of the hardware cores that can operate on higher frequencies to achieve higher throughputs. In addition, future developments comprise embedding of the infix processing step in hardware and widening the pool of implemented rules to increase extraction accuracy.